\documentclass[twocolumn,prl,amsmath,showpacs,amssymb,superscriptaddress,a4paper]{revtex4}

\usepackage{graphicx}
\usepackage{dcolumn}
\usepackage{bm}
\usepackage[sort&compress]{natbib}

\newcommand{\grad}{\ensuremath{^\circ}}
\newcommand*{\unit}[1]{\,\mathrm{#1}}

\newcommand{\coob}[1]{$\mathbf{CoO_{2}}$~}
\newcommand{\coo}[1]{$\mathrm{CoO_{2}}$~}
\newcommand{\naxb}[1]{$\mathbf{Na_{x}CoO_{2}}$~}
\newcommand{\nax}[1]{$\mathrm{Na_{x}CoO_{2}}$~}
\newcommand{\naab}[1]{$\mathbf{Na_{0.8}CoO_{2}}$~}
\newcommand{\naa}[1]{$\mathrm{Na_{0.8}CoO_{2}}$~}
\newcommand{\naas}[1]{$\mathrm{Na_{0.8}CoO_{2}}$}
\newcommand{\nabb}[1]{$\mathbf{Na_{0.85}CoO_{2}}$~}
\newcommand{\nabbs}[1]{$\mathbf{Na_{0.85}CoO_{2}}$}
\newcommand{\nab}[1]{$\mathrm{Na_{0.85}CoO_{2}}$~}
\newcommand{\nabs}[1]{$\mathrm{Na_{0.85}CoO_{2}}$}
\newcommand{\nac}[1]{$\mathrm{Na_{0.3}CoO_{2}}$~}
\newcommand{\nad}[1]{$\mathrm{Na_{0.7}CoO_{2}}$~}

\begin{document}

\title{Direct link between low-temperature magnetism\\ and high-temperature sodium order in $\mathbf{Na_{x}CoO_{2}}$}

\author{T.~F.~Schulze}
 \thanks{corresponding author, email: tschulze@phys.ethz.ch}
\author{P.~S.~H\"afliger}
\affiliation{Laboratory for Solid State Physics, ETH Z\"urich, CH-8093 Z\"urich, Switzerland}
\author{Ch.~Niedermayer}
 \affiliation{Laboratory for Neutron Scattering, ETH Z\"urich \& Paul Scherrer Institut (PSI), CH-5232 Villigen, Switzerland}
\author{K.~Mattenberger}
\author{S.~Bubenhofer}
\author{B.~Batlogg}
\affiliation{Laboratory for Solid State Physics, ETH Z\"urich, CH-8093 Z\"urich, Switzerland}

\date{\today}

\begin{abstract}
We prove the direct link between low temperature ($T$) magnetism and high-$T$ $\mathrm{Na^{+}}$ ordering in \nax\ using the example of a so far unreported magnetic transition at $8\unit{K}$ which involves a weak ferromagnetic moment. The $8\unit{K}$ feature is characterized in detail and its dependence on a diffusive $\mathrm{Na^{+}}$ rearrangement around $200\unit{K}$ is demonstrated. Applying muons as local probes this process is shown to result in a reversible phase separation into distinct magnetic phases that can be controlled by specific cooling protocols. Thus the impact of ordered $\mathrm{Na^{+}}$ Coulomb potential on the \coo\ physics is evidenced opening new ways to experimentally revisit the \nax\ phase diagram.
\end{abstract}


\maketitle

The layered transition metal oxide \nax\ combines high thermopower and metallic conductivity, thus being a promising candidate for thermoelectric energy conversion \cite{ter97,lee06}. Superconductivity \cite{tak03} in hydrated \nac\ and other types of ordered electronic groundstates for higher sodium content \cite{foo04,boo04,sal04,bay04,woo05,hel06} reflect the richness in its physics. The propensity of the \coo\ electronic system to a number of instabilities was revealed by theoretical studies assuming a homogeneous electronic structure \cite{Sin00,ind05}. However, the sodium ions (whose apparent main role is to donate electrons to the Co-O derived states) create local electronic inhomogeneity through their Coulomb potential, seen e.g.\ in NMR in the form of inequivalent Co sites \cite{muk05}. It was suggested from a theoretical point of view that sodium ordering at high temperatures \cite{zand04,hua04b,rog07} influences the electronic structure of the \coo\ layers through Coulomb potential wells \cite{rog07,geck06,zhan05}. The key experimental challenge in this field is to establish the link between high-$T$ sodium ordering and the low-$T$ magnetic properties. To verify such a link, control of sodium rearrangement would have to be achieved and the ensuing magnetic order to be monitored and quantified. Here we demonstrate this direct correspondance using the example of a new magnetic instability at $8\unit{K}$ in \naa\ and \nabs\ . We show it to result from diffusive sodium rearrangement in a narrow $T$ window near $200\unit{K}$, which is fully reversible and reproducible and can be completely controlled by the details of the samples' cooling protocol. This discovery reveals \nax\ to be the only system comprising itinerant electrons on a triangular lattice which is susceptible to an externally controllable Coulomb potential. It leads the way to experimental exploration of further sodium rearrangement processes through defined heat treatment protocols possibly yielding even more electronic instabilites in the same sample.

\begin{figure}
\includegraphics[width=0.42\textwidth]{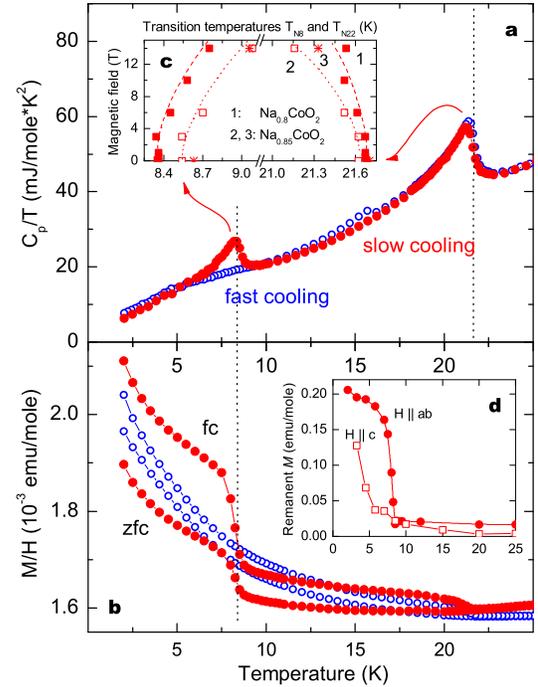}
\caption{\label{fig:masterplot} Presence of the $8\unit{K}$ magnetic phase in the 'slowly cooled' state, and absence of it in the 'fast cooled' state of the same \nab\ crystal as seen in specific heat (a) and magnetization (b, measured in field cooled (fc) and zero field cooled (zfc) mode for $H = 1 \unit{kOe}$ and $H~||~ab$). (c) The shift of transition temperatures in a magnetic field (data from $C_{p}$ measurements). 
(d) Remanent magnetization extracted from $M(H)$ loops up to $H = \pm 4\unit{kOe}$ for $H~||~ab$ and $H~||~c$.}
\end{figure}

Crystals were grown applying a standard preparation technique \cite{prab04} and analyzed using single crystal X-ray diffraction which generally confirmed the samples to be single-crystalline, the mosaic spread being $\leq1.5\grad$. Some samples showed twinning, but in all cases the c-axis and the ab-planes were well-defined. 
We begin with the description of the $8\unit{K}$ magnetic phase, which is completely dependent on the cooling procedure as is demonstrated in Fig.~\ref{fig:masterplot}. In the following, 'slow cooling' denotes a cooldown from ambient temperature with $\leq10\unit{K/min}$ while a 'fast cooled' sample has been directly inserted into the cryostat at $\leq100\unit{K}$. Fig.~\ref{fig:masterplot}a shows the specific heat in zero field (measured with relaxation technique on a Quantum Design PPMS-14) for both cooling protocols. In the 'slow cool' curve, in addition to an anomaly at $22\unit{K}$ there is another, lambda-shaped feature in $C_{p}$ clearly indicating a second phase transition at $8\unit{K}$. The corresponding entropies are approximately $60\unit{mJ/mole \cdot K}$ for the $22\unit{K}$ and $18\unit{mJ/mole \cdot K}$ for the $8\unit{K}$ peak. After 'fast cooling', the $8\unit{K}$ peak is absent while the anomaly at $22\unit{K}$ is larger by $\approx10\unit{\%}$. The coefficient $\gamma$ obtained from a fit to a two-phonon model between $25\unit{K}$ and $300\unit{K}$ is $18\unit{mJ/mole \cdot K^{2}}$ for both curves while the residual $\gamma$ is much smaller ($\gamma \leq5...7\unit{mJ/mole \cdot K^{2}}$ for $T\rightarrow0\unit{K}$).

The magnetic signature of the $8\unit{K}$ transition (measured on a Quantum Design MPMS XL SQUID magnetometer) is shown in Fig.~\ref{fig:masterplot}b, again for the two cooling protocols. In addition to the previously reported $22\unit{K}$ transition's features (e.g.~Refs.~\cite{sal04,bay04,woo05,mot03}), $M(T)$ measured with $H$ parallel to the \coo\ layers ($H~||~ab$) increases step-like at $8\unit{K}$ only for the 'slowly cooled' sample, suggesting the build up of a small ferromagnetic (FM) component. To clarify its orientation, the remanent magnetization was derived from $M(H)$ loops on a 'slowly cooled' sample with $H~||~ab$ and $H~||~c$ (Fig.~\ref{fig:masterplot}d). $M_{rem}$ sharply increases at $8.3 \unit{K}$ for $H~||~ab$ while no anomaly is seen for $H~||~c$, marking the development of a small FM component ($\approx 1 \cdot 10^{-4} \unit{\mu_{B}/mole~Co}$), oriented in the $ab$-plane. The FM character of the $8\unit{K}$ transition is also reflected in the shift of the transition temperature ($T_{N8}$) in an applied magnetic field up to $14\unit{T}$ (Fig.~\ref{fig:masterplot}c): The $22\unit{K}$ anomaly shifts to lower temperatures reflecting the onset of antiferromagnetic (AFM) order while the $8\unit{K}$ transition moves to higher temperatures by $0.4...0.5\unit{K}$, fortifying the idea of a FM transition. All presented measurements were repeated on $>15$ samples from the same and other batches yielding consistent results.

\begin{figure}
\includegraphics[width=0.42\textwidth]{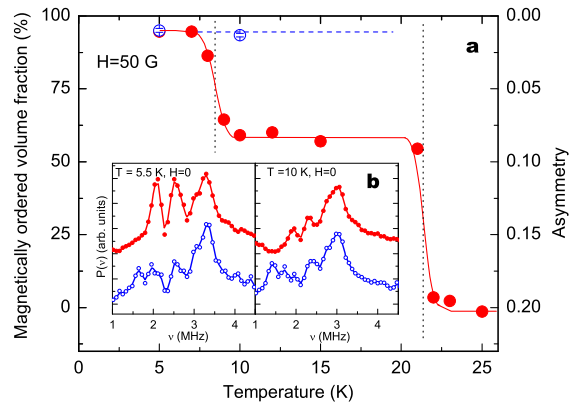}
\caption{\label{fig:usr} (a) Magnetically ordered volume fraction as determined from transverse-field $\mu SR$ data on \nab\ (Right scale: asymmetry of $0.676\unit{MHz}$ Gaussian contribution imposed by external field.) (b) Fourier spectra of asymmetry histograms for zero-field measurements give the spectral distribution of muon frequencies. 'slowly cooled': red, 'fast cooled': blue. Solid lines are a guide to the eye.}
\end{figure}

In the light of the small entropies associated with the anomalies in $C_{p}$ it is highly desirable to apply a technique that magnetically probes the bulk of the sample to determine the volume fractions involved in the two transitions.
To this end we performed Muon Spin Rotation ($\mu SR$) measurements at the Paul Scherrer Institut (PSI), Switzerland. We measured with and without external field in the 'slowly cooled' and 'fast cooled' state. If a small field ($50\unit{Oe}$ in our case) is applied transverse to the muon spin polarization, only the muons which come to rest in non-magnetic regions of the sample will precess with a frequency $\nu_{\mu}=B_{ext} \cdot 135.5\unit{MHz/T}$. Muons stopping in magnetically ordered regions experience a superposition of the internal and external field resulting in a drastically different time evolution of the muon spin polarization. This enables us to map the development upon cooling of the magnetically ordered volume fraction as shown in Fig.~\ref{fig:usr}a. It will be discussed in the following in the context of the internal magnetic fields seen in zero-field measurements, which provide a fingerprint for the identification of different magnetic phases. We obtained the dominant zero-field muon frequencies by fitting the time-domain spectra using standard procedures described elsewhere \cite{bay04,men05,sug03}. Here we restrict ourselves to showing the frequency distribution obtained by Fourier transform (Fig.~\ref{fig:usr}b).

\begin{figure*}[t]
\includegraphics[width=0.90\textwidth]{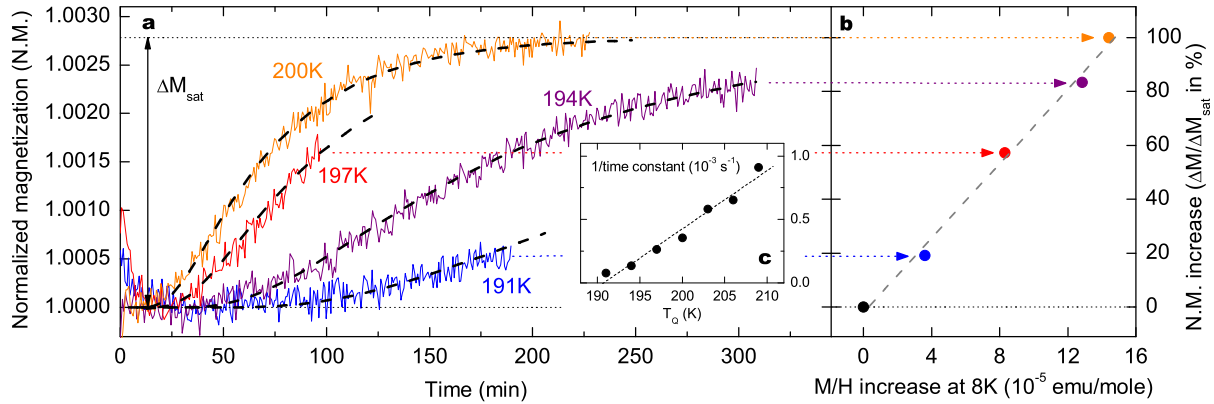}
\caption{\label{fig:mt} Magnetic signature of the Na rearrangement near $200\unit{K}$ in a \naa\ crystal. (a) Normalized magnetization as a function of time while sample is kept at $T_{Q}$ before subsequent cooldown to monitor the size of the $8\unit{K}$ transition. (b) Direct correspondance between the degree of Na rearrangement around $200\unit{K}$ and the FM moment developing at $8\unit{K}$ (measured as the increase in $M/H$). (c) Inverse time constant of diffusion fit to $M(t)$ (broken lines in (a)) vs. quenching temperature $T_{Q}$.}
\end{figure*}

The results of the transverse-field experiments imply that both cooling procedures result in a fully ordered sample volume at $5\unit{K}$. 
However, the microscopic magnetic state of the sample is distinctly different depending on the cooling protocol as implied by both the transverse and zero field data shown in Fig.~\ref{fig:usr}. The 'fast cooling' protocol results in a magnetic state that develops at $22\unit{K}$ and is closely similar to the state previously reported for 'fast cooled' $x=0.82$ samples \cite{bay04, bern07} (three dominant muon frequencies at $1.66(14)\unit{MHz}$, $2.39(10)\unit{MHz}$ and $2.93(10)\unit{MHz}$ in zero field at $10\unit{K}$). Within experimental limits the entire sample volume is in this particular state and it remains essentially unchanged upon cooling from $10\unit{K}$ to $5\unit{K}$ (Fig.~\ref{fig:usr}b).
When the very same crystal is 'slowly cooled', two distinct magnetic phases are sensed by the muons. The first one develops at $22\unit{K}$ and involves approximately $65\unit{\%}$ of the sample volume. The details of the frequency distribution at $10\unit{K}$ in zero field are slightly different from the 'fast cooled' $22\unit{K}$ phase ($1.95(15)\unit{MHz}$, $2.33(15)\unit{MHz}$ and a dominant $2.90(7)\unit{MHz}$ signal). In addition, the spectrum contains a significant Kubo-Toyabe contribution whose weight is consistent with the paramagnetic volume fraction as obtained in the transverse field measurements. The comparison with previous detailed $\mu SR$ studies on apperently 'fast cooled' samples \cite{bern07} implies a Na content of $x=0.75...0.8$ for this phase. Below $8\unit{K}$, magnetic order develops in the remaining $\approx 35\unit{\%}$ of the sample volume causing a prominent change of the zero-field frequency distribution (Fig.~\ref{fig:usr}b). Thus, the data suggest two distinct magnetically ordered phases to coexist at $5\unit{K}$ in the 'slowly cooled' state with different Na content while a single magnetically ordered phase is observed in 'fast cooled' samples. 

We continue by analyzing the dependence of the $8\unit{K}$ magnetic feature on the cooling protocol. This is done by studying the magnetization as the magnetometer allows for convenient shock cooling. We minimized the equilibration time by choosing a small $6.3 \unit{mg}$ single crystal. A quenching cycle consisted of the following steps: a) Holding the sample at room temperature ($296\unit{K}$) for at least 20 minutes. b) Quenching it to a preset temperature ($T_{Q}$). c) Further cooling at $10\unit{K/min}$ and measuring $M(T)$ between $25\unit{K}$ and $2\unit{K}$ to monitor the development of the $8\unit{K}$ magnetic phase.

A first essential observation was that quenching to any temperature below $212\unit{K}$ completely inhibits the $8\unit{K}$ transition while the $22\unit{K}$ transition is unaffected, but setting $T_{Q}$ to $215\unit{K}$ yields a fully developed $8\unit{K}$ transition. For $T_{Q}=191...212\unit{K}$, waiting at $T_{Q}$ before successive further cooldown enables the $8\unit{K}$ transition. Obviously, a distinct change happening in this narrow temperature range causes the formation of the $8\unit{K}$ magnetic phase.
As we noticed a minute increase of the magnetization associated with this transition, we could follow its development in time. This was done by quenching the sample to and keeping it at $T_{Q}$ while measuring $M$ over an extended period of time.
The time evolution for a few selected $T_{Q}$ is shown in Fig.~\ref{fig:mt}a, the $200\unit{K}$ curve being an example of a complete time evolution. We followed $M(t)$ for $T_{Q}=190...209\unit{K}$ and observed a universal relative increase by $0.0028\cdot M_{initial}$. The lower $T_{Q}$ is, the more time is needed for saturation. If the phase development is interrupted by further rapid cooling, the resulting $8\unit{K}$ FM component is smaller, indicating a smaller amount of $8\unit{K}$ magnetic phase. Fig.~\ref{fig:mt}b shows the one-to-one correspondence between the degree of the transition around $200\unit{K}$ and the amount of resulting $8\unit{K}$ phase. Apparently, the $8\unit{K}$ phase grows with time by a diffusive process as suggested by the close agreement between the $M(t)$ data and the broken lines in Fig.~\ref{fig:mt}a, that represent solutions to the non-stationary diffusion equation $\Delta c \propto \tau \cdot(\partial c/\partial t)$ with the time constant $\tau$ being the only parameter. This time constant increases more then tenfold upon cooling in the $20\unit{K}$ interval studied (Fig.~\ref{fig:mt}c). The shock cooling experiments were repeated $>20$ times on the same sample with excellent reproducibility. Other samples yielded compatible results with $\Delta M_{sat}$ being sample-dependent.

We infer that \textit{diffusive sodium rearrangement} near $200\unit{K}$ leads to the formation of the $8\unit{K}$ magnetic phase.
Keeping in mind that the magnetic interplane coupling $J_{c}$ is determined by superexchange processes involving Na orbitals \cite{joh05} it is consistent that a signature of ongoing Na rearrangement can be seen in the magnetization. The freezing of the time constant reflects the complex interplay between the intrinsic propensity to Na ordering and the temperature dependent mobility of the Na ions. It is desirable to directly probe the Na ions to confirm its dynamics. Indeed, in recent NMR studies on \nad\ powder, various anomalies in the Na NMR spectrum were seen in a similar $T$ range \cite{gav05}.

The result of the rearrangement is the coexistence of two distinct phases with different sodium content as revealed by $\mu SR$.
Knowing the volume ratio of $65:35$ for the $x=0.85$ crystals and $x=0.75...0.8$ for the stable $22\unit{K}$ phase from $\mu SR$ we conclude that $x=0.95...1$ for the stable $8\unit{K}$ phase. Phase separations have been invoked before to explain either the coexistence of multiple crystallographic phases or the presence of paramagnetic volume in otherwise magnetically ordered samples \cite{hua04,devau05,sak04}. Here the actual process of phase separation is followed and the direct link to low temperature electronic instabilities is proven. Whether the driving force of the separation lies within the Na dynamics itself or is controlled by an instability of the \coo\ electronic system towards charge ordering which in turn triggers the Na rearrangement is an interesting open question.

\begin{figure}[t]
\includegraphics[width=0.42\textwidth]{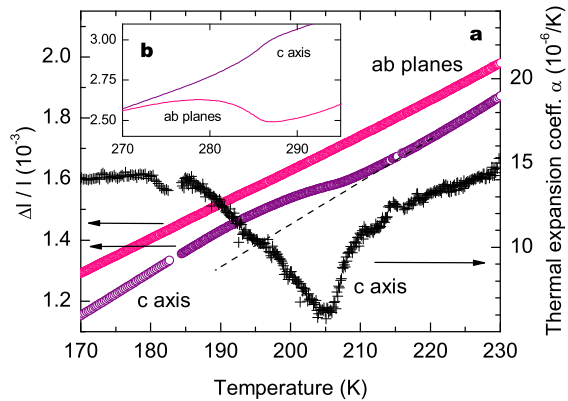}
\caption{\label{fig:thexp} Lattice distortion caused by Na rearrangement. Thermal expansion anomalies in a \nab\ crystal due to the Na rearrangement near $200\unit{K}$ (a) and Na ordering transition at $285\unit{K}$ (b), $ab$-plane data multiplied by 2.85.}
\end{figure}

The Na rearrangement process around $200\unit{K}$ is remarkebly different from the well-documented $285\unit{K}$ Na ordering transition \cite{rog07}. While we observe a specific heat anomaly at $285\unit{K}$, we do not detect a $C_{p}$ signature around $200\unit{K}$. However, using the high sensitivity of thermal expansion measurements (applying a miniature capacitive dilatometer developed at the TU Vienna \cite{rot98}) we did find a pronounced signal: The crystal expands by more than $1.3 \cdot 10^{-4}$ along the $c$-axis upon cooling between $210\unit{K}$ and $190\unit{K}$. We note that no feature is seen in the $ab$-plane (Fig.~\ref{fig:thexp}a). A sharp lambda-like anomaly marks the Na rearrangement in the c-axis' thermal expansion coefficient. For comparison we show in Fig.~\ref{fig:thexp}b the expansion signal of the $285\unit{K}$ Na ordering in the same sample.

In conclusion we present evidence of a new magnetic transition below $8\unit{K}$ in \nax\ ($x$=0.80 and 0.85) involving a small ferromagnetic in-plane component. It is associated with the formation of two distinct magnetic phases of different Na content as implied by $\mu$SR studies. The $8\unit{K}$ phase is stabilized by sufficiently slow cooling through the narrow temperature range around $200\unit{K}$ where diffusive Na rearrangement takes place. Thus the link between high-$T$ Na rearrangement and low-$T$ magnetic properties is established.

In analogy to other ionic ordering phenomena one could expect a hierarchy of successive rearrangement processes to happen, resulting in more than the two phases discussed above.
In this light, the apparently unreconcilable reports on different AFM transition temperatures for the same nominal Na content (e.g. $27\unit{K}$ \cite{men05}, $22\unit{K}$ \cite{sak04}, $19.8\unit{K}$ \cite{bay04}, and $18.5\unit{K}$ \cite{luo04} for $x=0.82...0.85$) may well be the result of a variety of Na ordering patterns and their associated magnetic transitions.
Having connected high-$T$ Na ordering and low-$T$ magnetism, the challenge will be twofold: To experimentally revisit the phase diagram \cite{foo04,yos07b} making use of Na arrangement as an additional control parameter and further, in the broad context of correlated electron physics, to theoretically explore the interplay between narrow band itinerant electron behaviour and localization induced by the patterned Na potential.

We would like to thank H. M\"uller and M. Rotter for their help with the thermal expansion measurements and S. Katrych for performing XRD. This work is partly based on muon experiments performed at the Swiss Muon Source, Paul Scherrer Institut, Villigen (Switzerland).


\bibliographystyle{apsrev}

\end{document}